\documentclass[12pt,english]{article}
\usepackage[T1]{fontenc}
\usepackage[latin9]{inputenc}
\usepackage[a4paper]{geometry}
\usepackage{amsmath}
\usepackage{setspace}
\usepackage{amssymb}
\makeatletter
\newcommand{\lyxaddress}[1]{
\par {\raggedright #1
\vspace{1.4em}
\noindent\par}
}

\makeatother

\usepackage{babel}

\begin{document}

\title{GENERALIZED SPLIT-OCTONION ELECTRODYNAMICS}

\author{B. C. Chanyal, P. S. Bisht and O. P. S. Negi%
\thanks{Present Address from November 08- December22, 2010:\textbf{ Universität
Konstanz, Fachbereich Physik, Postfach-M 677, D-78457 Konstanz, Germany}%
}}

\maketitle
\begin{singlespace}

\lyxaddress{\begin{center}
Department of Physics\\
Kumaun University\\
S. S. J. Campus\\
 Almora -263601 (U.K.) India
\par\end{center}}
\end{singlespace}

\lyxaddress{\begin{center}
Email:- bcchanyal@gmail.com\\
ps\_bisht123@rediffmail.com\\
ops\_negi@yahoo.co.in 
\par\end{center}}
\begin{abstract}
Starting with the usual definitions of octonions and split octonions
in terms of Zorn vector matrix realization, we have made an attempt
to write the consistent form of generalized Maxwell\textquoteright{}s
equations in presence of electric and magnetic charges (dyons). We
have thus written the generalized potential, generalized field, and
generalized current of dyons in terms of split octonions and accordingly
the split octonion forms of generalized Dirac Maxwell\textquoteright{}s
equations are obtained in compact and consistent manner. This theory
reproduces the dynamic of electric (magnetic) in the absence of magnetic
(electric) charges. 


\end{abstract}

\section{Introduction}

~~~~~~~The relationship between mathematics and physics has
long been an area of interest and speculation. So, there has been
a revival in the formulation of natural laws so that there exists
\cite{key-1} four-division algebras consisting the algebra of real
numbers ($\mathbb{R}$), complex numbers ($\mathbb{C}$), quaternions
($\mathbb{H}$) and Octonions ($\mathcal{O}$). All four algebra's
are alternative with totally anti symmetric associators. Quaternions
\cite{key-2,key-3} were very first example of hyper complex numbers
have been widely used \cite{key-4,key-5,key-6,key-7,key-8,key-9,key-10}
to the various applications of mathematics and physics. Since octonions
\cite{key-11} share with complex numbers and quaternions, many attractive
mathematical properties, one might except that they would be equally
as useful as others. Octonion \cite{key-11} analysis has been widely
discussed by Baez \cite{key-12}. It has also played an important
role in the context of various physical problems \cite{key-13,key-14,key-15,key-16}
of higher dimensional supersymmetry, super gravity and super strings
etc. In recent years, it has also drawn interests of many \cite{key-17,key-18,key-19,key-20}
towards the developments of wave equation and octonion form of Maxwell's
equations. We \cite{key-21,key-22} have also studied octonion electrodynamics,
dyonic field equation and octonion gauge analyticity of dyons consistently
and obtained the corresponding field equations (Maxwell's equations)
and equation of motion in compact and simpler formulation. Keeping
in view the recent interests on the existence of monoploes and dyons
at one end and quaternion-octonion formulation of generalized Dirac-Maxwell's
(GDM) equations at oher end, we \cite{key-23} have reformulated the
generalized Dirac-Maxwell's equations of dyons by means of octonion
variables in compact and consistent manner. Starting with the usual
definitions of octonions and split octonions in terms of Zorn vector
matrix realization, in this paper, we have made an attempt to extend
the present study to write the consistent form of generalized Dirac-Maxwell\textquoteright{}s
(GDM) equations of dyons in terms of split octonion variables without
imposing constraints on physical variables as we \cite{key-23} did
for the case of normal octonions. We have thus written the generalized
potential, generalized field, and generalized current of dyons in
terms of split octonions and accordingly the split octonion forms
of generalized Dirac Maxwell\textquoteright{}s equations are derived
in compact and consistent manner. This theory reproduces the dynamic
of electric (magnetic) in the absence of magnetic (electric) charges.

\section{Octonion Definition }

An octonion $x$ is expressed \cite{key-21,key-22} as a set of eight
real numbers

\begin{eqnarray}
x=(x_{0},\, x_{1},....,\, x_{7}) & = & x_{0}e_{0}+\sum_{A=1}^{7}x_{A}e_{A}\,\,\,\,\,\,\,\,\,\,\,(A=1,2,.....,7)\label{eq:1}\end{eqnarray}
where $e_{A}(A=1,2,.....,7)$ are imaginary octonion units and $e_{0}$
is the multiplicative unit element. The octet $(e_{0},e_{1},e_{2},e_{3},e_{4},e_{5},e_{6},e_{7})$
is known as the octonion basis and its elements satisfy the following
multiplication rules

\begin{eqnarray}
e_{0}=1,\,\,\,\, & e_{0}e_{A}=e_{A}e_{0}=e_{A}\,\,\, & e_{A}e_{B}=-\delta_{AB}e_{0}+f_{ABC}\, e_{C}.\,\,\,\,\,(A,B,C=1,2,......7)\label{eq:2}\end{eqnarray}
The structure constants $f_{ABC}$ are completely antisymmetric and
take the value $1$ i.e.

$f_{ABC}=+1=(123),\,(471),\,(257),\,(165),\,(624),\,(543),\,(736)$.

Here the octonion algebra $\mathcal{O}$ is described over the algebra
of rational numbers having the vector space of dimension $8$. Octonion
algebra is non associative. Octonion conjugate is thus defined as,

\begin{eqnarray}
\bar{x} & = & x_{0}e_{0}-\sum_{A=1}^{7}x_{A}e_{A}\,\,\,\,\,\,\,\,\,\,\,(A=1,2,.....,7).\label{eq:3}\end{eqnarray}
The norm of the octonion $N(x)$ is defined as

\begin{eqnarray}
N(x)=\overline{x}x & =x\,\bar{x} & =\sum_{\alpha=0}^{7}x_{\alpha}^{2}e_{0}\label{eq:4}\end{eqnarray}
which is zero if $x=0$, and is always positive otherwise. It also
satisfies the following property of normed algebra

\begin{eqnarray}
N(xy) & =N(x)N(y) & =N(y)N(x).\label{eq:5}\end{eqnarray}
As such, for a nonzero octonion $x$ , we may define its inverse as
$x^{-1}=\frac{\bar{x}}{N(x)}$which shows that $x^{-1}x=xx^{-1}=1.e_{0};\,\,\,\,(xy)^{-1}=y^{-1}x^{-1}.$

\section{Split Octonions}

The split octonions are a non assosiative extension of split quaternions.
.They differ from the octonion in the signature of quadratic form.
the split octonions have a signature $(4,4)$ whereas the octonions
have positive signature $(8,0)$. The Cayley algebra of octonions
over the field of complex numbers $\mathbb{C_{C}=C\otimes}C$ is visualized
as the algebra of split octonions with its following basis elements,

\begin{eqnarray}
u_{0}=\frac{1}{2} & (1+i\,\, e_{7}),\,\,\,\,\,\,\,\,\,\,\, & u_{0}^{\star}=\frac{1}{2}(1-i\,\, e_{7}),\nonumber \\
u_{1}=\frac{1}{2} & (e_{1}+i\,\, e_{4}),\,\,\,\,\,\,\,\,\,\,\, & u_{1}^{\star}=\frac{1}{2}(e_{1}-i\,\, e_{4}),\nonumber \\
u_{2}=\frac{1}{2} & (e_{2}+i\,\, e_{5}),\,\,\,\,\,\,\,\,\,\,\, & u_{2}^{\star}=\frac{1}{2}(e_{2}-i\,\, e_{5}),\nonumber \\
u_{3}=\frac{1}{2} & (e_{3}+i\,\, e_{6}),\,\,\,\,\,\,\,\,\,\,\, & u_{3}^{\star}=\frac{1}{2}(e_{3}-i\,\, e_{6}),\label{eq:6}\end{eqnarray}
where $(\star)$ is used for complex conjugation for which $(i=\sqrt{-1})$
is usual complex imaginary number and commutes with all the seven
octonion imaginary units $e_{A}(A=1,2...,7)$. The split octonion
basis elements satisfy the following multiplication rules \cite{key-13};

\begin{eqnarray}
u_{i}u_{j} & = & \epsilon_{ijk}u_{k}^{\star};\,\,\,\,\, u_{i}^{\star}u_{j}^{\star}=-\epsilon_{ijk}u_{k}(\forall i,j,k=1,2,3)\nonumber \\
u_{i}u_{j}^{\star} & = & -\delta_{ij}u_{0};\,\,\,\,\, u_{i}u_{0}=0;\,\,\,\,\, u_{i}^{\star}u_{0}=u_{i}^{\star}\nonumber \\
u_{i}^{\star}u_{j} & = & -\delta_{ij}u_{0};\,\,\,\,\,\, u_{i}u_{0}^{\star}=u_{0};\,\,\,\,\, u_{i}^{\star}u_{0}^{\star}=0\nonumber \\
u_{0}u_{i} & = & u_{i};\,\,\,\,\, u_{0}^{\star}u_{i}=0;\,\,\,\,\,\, u_{0}u_{i}^{\star}=0\nonumber \\
u_{0}^{\star}u_{i}^{\star} & = & u_{i};u_{0}^{2}=u_{0};\,\,\,\,\, u_{0}^{\star2}=u_{0}^{\star};\,\,\,\,\, u_{0}u_{0}^{\star}=u_{0}^{\star}u_{0}=0\label{eq:7}\end{eqnarray}
as the bi-valued representations of quaternion units $\begin{array}{c}
e_{0}\end{array},\: e_{1},\: e_{2},\: e_{3}$. We may thus introduce a convenient realization for the basis elements
$(u_{0},u_{i},u_{0}^{\star},u_{i}^{\star})$ in terms of Pauli spin
matrices as

\begin{eqnarray}
u_{0} & = & \left[\begin{array}{cc}
0 & 0\\
0 & 1\end{array}\right];\,\,\,\,\,\,\,\,\,\, u_{0}^{\star}=\left[\begin{array}{cc}
1 & 0\\
0 & 0\end{array}\right];\nonumber \\
u_{i} & = & \left[\begin{array}{cc}
0 & 0\\
e_{j} & 0\end{array}\right];\,\,\,\,\,\,\,\,\,\, u_{j}=\left[\begin{array}{cc}
0 & -e_{j}\\
0 & 0\end{array}\right](\forall j=1,2,3)\label{eq:8}\end{eqnarray}
where $1,\, e_{1},\, e_{2},\, e_{3}$ are quaternion units satisfying
the multiplication rule $e_{j}e_{k}=-\delta_{jk}+\epsilon_{jkl}e_{l}$.
The split Caley octonion algebra is thus expressed in terms of 2$\times$2
Zorn's vector matrix realizations as

\begin{align}
A=au_{0}^{*}+bu_{0}+x_{j}u_{j}^{*}+y_{j}u_{j}= & \left(\begin{array}{cc}
a & -\overrightarrow{x}\\
\overrightarrow{y} & b\end{array}\right)\label{eq:9}\end{align}
 The split octonion conjugation of equation (\ref{eq:9}) is then
described as,

\begin{eqnarray}
\overline{A} & =au_{0}+bu_{0}^{\star}-x_{j}u_{j}^{\star}-y_{j}u_{j} & =\left(\begin{array}{cc}
b & \overrightarrow{x}\\
-\overrightarrow{y} & a\end{array}\right).\label{eq:10}\end{eqnarray}
The norm of $A$ is defined as $\overline{A}A=A\overline{A}=(ab+\overrightarrow{x}.\overrightarrow{y})\hat{1}$
with $\hat{1}$ as the unit matrix of order $2\times2$. Any four
- vector $A_{\mu}$ (complex or real) can equivalently be written
in terms of the following Zorn matrix realization as 

\begin{eqnarray}
Z(A) & = & \left(\begin{array}{cc}
x_{4} & -\overrightarrow{x}\\
\overrightarrow{y} & y_{4}\end{array}\right);\,\,\,\, Z(\overline{A})=\left(\begin{array}{cc}
x_{4} & \overrightarrow{x}\\
-\overrightarrow{y} & y_{4}\end{array}\right).\label{eq:11}\end{eqnarray}
So we may write the split octonion differential operator $\boxdot$
and its conjugate $\overline{\boxdot}$ in terms of the $2\times2$
Zorn matrix realization as

\begin{eqnarray}
\boxdot & = & \left(\begin{array}{cc}
\partial_{t} & -\overrightarrow{\nabla}\\
\overrightarrow{\nabla} & -\partial_{t}\end{array}\right);\,\,\,\,\,\,\,\,\,\,\overline{\boxdot}=\left(\begin{array}{cc}
-\partial_{t} & \overrightarrow{\nabla}\\
-\overrightarrow{\nabla} & \partial_{t}\end{array}\right),\label{eq:12}\end{eqnarray}
where $\partial_{t}=\frac{\partial}{\partial t}$. As such , we get

\begin{alignat}{1}
\boxdot & \overline{\boxdot}=\left(\begin{array}{cc}
\nabla^{2}-\frac{\partial^{2}}{\partial t^{2}} & 0\\
0 & \nabla^{2}-\frac{\partial^{2}}{\partial t^{2}}\end{array}\right)=\overline{\boxdot}\boxdot=\square,\label{eq:13}\end{alignat}
where $\begin{array}{c}
\nabla^{2}=\frac{\partial^{2}}{\partial x^{2}}+\frac{\partial^{2}}{\partial y^{2}}+\frac{\partial^{2}}{\partial z^{2}}\end{array}$and $\begin{array}{c}
\mathtt{\square}=\frac{\partial^{2}}{\partial x^{2}}+\frac{\partial^{2}}{\partial y^{2}}+\frac{\partial^{2}}{\partial z^{2}}-\frac{\partial^{2}}{\partial t^{2}}=\nabla^{2}-\frac{\partial^{2}}{\partial t^{2}}\end{array}$.

\section{Generalized Split Octonion Electrodynamics}

Let us start with octonion form of generalized potential \cite{key-22}
of dyons as 

\begin{alignat}{1}
\mathbb{V} & =e_{0}V_{0}+e_{1}V_{1}+e_{2}V_{2}+e_{3}V_{3}+e_{4}V_{4}+e_{5}V_{5}+e_{6}V_{6}+e_{7}V_{7}.\label{eq:14}\end{alignat}
Here we have described the components of generalized four potential
of dyons as

$(V_{0},V_{1},V_{2},V_{3},V_{4},V_{5},V_{6},V_{7})\Longrightarrow(\varphi,\, A_{x},\, A_{y},A_{z},\, iB_{x},\, iB_{y},\, iB_{z},i\phi)\,\,\,\,(i=\sqrt{-1})$

with $(\phi,\, A_{x},\, A_{y},\, A_{z})=(\phi,\overrightarrow{A}\,)=\left\{ A_{\mu}\right\} $
and $(\varphi,\, B_{x},\, B_{y},\, B_{z})=(\varphi,\overrightarrow{B}\,)=\left\{ B_{\mu}\right\} $
are respectively described as the components of electric $\left\{ A_{\mu}\right\} $
and magnetic $\left\{ B_{\mu}\right\} $ four potential constituents
of dyons (particles carrying simultaneously the electric and magnetic
charges). Equation (\ref{eq:14}) may then be written \cite{key-23}
as 

\begin{align}
\mathbb{V=} & e_{1}(A_{x}+ie_{7}B_{x})+e_{2}(A_{y}+ie_{7}B_{y})+e_{3}(A_{z}+ie_{7}B_{z})+(\varphi+ie_{7}\phi)\nonumber \\
= & e_{1}\mathrm{V}_{\mathrm{x}}+e_{2}\mathrm{V}_{\mathrm{y}}+e_{3}\mathrm{V_{z}}+ie_{7}\emptyset.\label{eq:15}\end{align}
So, the split octonion form of generalized four potential of dyons
may be written in terms of 2$\times$2 Zorn's vector matrix realization
as

\begin{align}
\mathbb{V}= & \left(\begin{array}{cc}
\Phi_{-} & -\overrightarrow{V_{+}}\\
\overrightarrow{V_{-}} & \textrm{\ensuremath{\Phi}}_{+}\end{array}\right)=\left(\begin{array}{cc}
\left(\varphi-\phi\right) & -\left(\overrightarrow{A}+\overrightarrow{B}\right)\\
\left(\overrightarrow{A}-\overrightarrow{B}\right) & \left(\varphi+\phi\right)\end{array}\right).\label{eq:16}\end{align}
Here $\left[\Phi_{-}=\left(\varphi-\phi\right),\;\Phi_{+}=\left(\varphi+\phi\right),\;\overrightarrow{V_{-}}\rightarrow\left(\overrightarrow{A}-\overrightarrow{B}\right)\;\overrightarrow{V_{+}}\rightarrow\left(\overrightarrow{A}+\overrightarrow{B}\right)\right]$
with $\overrightarrow{A}=(A_{x},\, A_{y},\, A_{z})$ and $\overrightarrow{B}=(B_{x},\, B_{y},\, B_{z}).$
Now operating $\overline{\boxdot}$ (\ref{eq:12}) to octonion potential
$\mathtt{\mathrm{\mathcal{\mathsf{\mathbf{\mathcal{\mathfrak{\mathbb{V}}}}}}}}$
(\ref{eq:16}) , we get

\begin{align}
\overline{\boxdot}\mathbb{V=} & \left(\begin{array}{cc}
-\frac{\partial\varphi}{\partial t}+\frac{\partial\phi}{\partial t}+\overrightarrow{\nabla}.\overrightarrow{A}-\overrightarrow{\nabla}.\overrightarrow{B} & \frac{\partial\overrightarrow{A}}{\partial t}+\frac{\partial\overrightarrow{B}}{\partial t}+\overrightarrow{\nabla}\varphi+\overrightarrow{\nabla}\phi-\overrightarrow{\nabla}\times\overrightarrow{A}+\overrightarrow{\nabla}\times\overrightarrow{B}\\
-\overrightarrow{\nabla}\varphi+\overrightarrow{\nabla}\phi+\frac{\partial\overrightarrow{A}}{\partial t}-\frac{\partial\overrightarrow{B}}{\partial t}+\overrightarrow{\nabla}\times\overrightarrow{A}+\overrightarrow{\nabla}\times\overrightarrow{B} & \overrightarrow{\nabla}.\overrightarrow{A}+\overrightarrow{\nabla}.\overrightarrow{B}+\frac{\partial\varphi}{\partial t}+\frac{\partial\phi}{\partial t}\end{array}\right).\label{eq:17}\end{align}
It is to be noted that we have used S.I. system of natural units $(c=\hbar=1)$
through out the text. Equation (\ref{eq:14}) then reduces to \begin{align}
\overline{\boxdot}\,\mathbb{V} & =\mathbb{F};\label{eq:18}\end{align}
where $\mathbb{F}$ is also an octonion describing the generalized
electromagnetic fields of dyons \cite{key-23} as 

\begin{align}
\mathbb{F} & =\sum_{a=0}^{a=7}e_{a}F_{a}=e_{1}F_{1}+e_{2}F_{2}+e_{3}F_{3}+e_{4}F_{4}+e_{5}F_{5}+e_{6}F_{6}\nonumber \\
= & e_{1}(H_{x}+ie_{7}E_{x})+e_{2}(H_{y}+ie_{7}E_{y})+e_{3}(H_{z}+ie_{7}E_{z})\label{eq:19}\end{align}
with the componets $\begin{array}{c}
F_{0}=F_{7}=0\end{array}$ due to Lorentz gauge conditions applied for electric and magnetic
four potentials. Equation (\ref{eq:19}) may then be described as
a split octonion written in terms of 2$\times$2 Zorn's vector matrix
realization as

\begin{align}
\mathbb{F}= & \left(\begin{array}{cc}
0 & -\overrightarrow{\digamma_{+}}\\
\overrightarrow{\digamma_{-}} & 0\end{array}\right)=\left(\begin{array}{cc}
0 & -\left(\overrightarrow{\digamma_{g}}+\overrightarrow{\digamma_{e}}\right)\\
\left(\overrightarrow{\digamma_{g}}-\overrightarrow{\digamma_{e}}\right) & 0\end{array}\right)\label{eq:20}\end{align}
where $\begin{array}{c}
\overrightarrow{\digamma_{+}}=\overrightarrow{\digamma_{g}}+\overrightarrow{\digamma_{e}}\end{array}$,$\begin{array}{c}
\overrightarrow{\digamma_{-}}=\overrightarrow{\digamma_{g}}-\overrightarrow{\digamma_{e}}\end{array}$ and 

\begin{align}
\overrightarrow{\digamma_{g}}= & -\frac{\partial\overrightarrow{B}}{\partial t}-\overrightarrow{\nabla}\varphi+\overrightarrow{\nabla}\times\overrightarrow{A}.\;\longrightarrow\overrightarrow{H};\nonumber \\
\overrightarrow{\digamma_{e}}= & -\frac{\partial\overrightarrow{A}}{\partial t}-\overrightarrow{\nabla_{n}}\phi-\overrightarrow{\nabla_{n}}\times\overrightarrow{B_{n}};\;\longrightarrow\overrightarrow{E}.\label{eq:21}\end{align}
Here $\overrightarrow{E}$ and $\overrightarrow{H}$ are respectively
described as the generalized electric and magnetic fields of dyons
\cite{key-22}. As such, we may write the generalized electromagnetic
field vector $\mathbb{F}$ of dyons in terms of following split octonionic
representation i.e.

\begin{align}
\mathbb{F}= & \left(\begin{array}{cc}
0 & -\left(\overrightarrow{H}+\overrightarrow{E}\right)\\
\overrightarrow{H}-\overrightarrow{E} & 0\end{array}\right)=\left(\begin{array}{cc}
0 & +\overrightarrow{\psi_{+}}\\
\overrightarrow{\psi_{-}} & 0\end{array}\right)\label{eq:22}\end{align}
where $\overrightarrow{\psi_{+}}=\overrightarrow{H}+\overrightarrow{E}$
and $\begin{array}{c}
\overrightarrow{\psi_{-}}\end{array}=\overrightarrow{H}-\overrightarrow{E}$ are described the generalized electromagnetic vector fields of dyons.
Now operating the differential operator $\begin{array}{c}
\boxdot\end{array}$(\ref{eq:12}) to generalized vector field $\begin{array}{c}
\mathbb{F}\end{array}$ (\ref{eq:22}), we get 

\begin{align}
\boxdot\,\mathbb{\mathbb{F}=} & \left(\begin{array}{cc}
\overrightarrow{\nabla.}\overrightarrow{\digamma_{g}}-\overrightarrow{\nabla.}\overrightarrow{\digamma_{e}} & \frac{\partial\overrightarrow{\digamma_{g}}}{\partial t}+\frac{\partial\overrightarrow{\digamma_{e}}}{\partial t}-\overrightarrow{\nabla}\times\overrightarrow{\digamma_{g}}+\overrightarrow{\nabla}\times\overrightarrow{\digamma_{e}}\\
\frac{\partial\overrightarrow{\digamma_{g}}}{\partial t}-\frac{\partial\overrightarrow{\digamma_{e}}}{\partial t}+\overrightarrow{\nabla}\times\overrightarrow{\digamma_{g}}+\overrightarrow{\nabla}\times\overrightarrow{\digamma_{e}} & \overrightarrow{\nabla.}\overrightarrow{\digamma_{g}}+\overrightarrow{\nabla.}\overrightarrow{\digamma_{e}}\end{array}\right)\label{eq:23}\end{align}
which may also be written in terms of the following wave equation
in split octonion form as .

\begin{align}
\boxdot\mathbb{F} & =-\mathbb{J}.\label{eq:24}\end{align}
Here $\mathbb{J}$ is also an octonion which may be identified as
the octonion form of generalized four-current of dyons as

\begin{align}
\mathbb{J}= & \sum_{a=0}^{a=7}e_{a}\mathrm{j_{a}=\mathrm{e_{1}}(\mathrm{\mathrm{j}}_{\mathrm{x}}+i\mathrm{e_{7}k_{x})+\mathrm{e_{2}(j_{y}+ie_{7}k_{z})+e_{3}(j_{z}+k_{z})+(\varrho+ie_{7}\rho)}}.}\label{eq:25}\end{align}
It may also be written as the split octonion representation in terms
of 2$\times$2 Zorn's vector matrix realization as

\begin{align}
\mathbb{J}= & \left(\begin{array}{cc}
\left(\varrho-\rho\right) & -\left(\overrightarrow{j}+\overrightarrow{k}\right)\\
\left(\overrightarrow{j}-\overrightarrow{k}\right) & \left(\varrho+\rho\right)\end{array}\right)=\left(\begin{array}{cc}
\jmath_{-} & -\vec{\jmath}_{+}\\
\vec{\jmath}_{-} & \jmath_{+}\end{array}\right)\label{eq:26}\end{align}
where $\begin{array}{c}
\jmath_{-}\end{array}=\left(\varrho-\rho\right),\;\vec{\jmath}_{-}=\left(\overrightarrow{j}-\overrightarrow{k}\right),\;\jmath_{+}=\left(\varrho+\rho\right),\;\vec{\jmath_{+}}=\left(\overrightarrow{j}+\overrightarrow{k}\right).$ Here $\mathrm{(\rho,\,\overrightarrow{j})=\{\mathrm{j_{\mu}}\}}$,
$(\varrho,\,\overrightarrow{j})=\{\mathrm{k_{\mu}}\}$ and $(\mathrm{J_{0},\overrightarrow{J})=\{\mathrm{J_{\mu}}\}}$
are respectively the four currents associated with electric charge,
magnetic monopole and generalized fields of dyons. Equations (\ref{eq:24})
thus leads to the generalized Dirac Maxwell's (GDM) equations of dyons
\cite{key-22}. Using equations (\ref{eq:12},\ref{eq:13} and \ref{eq:26}),
we get 

\begin{align}
\boxdot\overline{\boxdot}\mathbb{V}= & \left(\begin{array}{cc}
\left(\square\varphi-\square\phi\right) & -\left(\square\overrightarrow{A}+\square\overrightarrow{B}\right)\\
\left(\square\overrightarrow{A}-\square\overrightarrow{B}\right) & \left(\square\varphi+\square\phi\right)\end{array}\right)=\left(\begin{array}{cc}
\left(\varrho-\rho\right) & -\left(\overrightarrow{j}+\overrightarrow{k}\right)\\
\left(\overrightarrow{j}-\overrightarrow{k}\right) & \left(\varrho+\rho\right)\end{array}\right)=\mathbb{J}\label{eq:27}\end{align}
which is described as the split potential wave equation for generalized
fields of dyons. Now operating $\overline{\boxdot}$ (\ref{eq:12})
to split-octonion current $\mathtt{\mathrm{\mathcal{\mathsf{\mathbf{\mathcal{\mathbb{J}}}}}}}$
(\ref{eq:26}), we get 

\begin{align}
\overline{\boxdot}\mathcal{\mathbb{J}}= & \left(\begin{array}{cc}
-\frac{\partial\varrho}{\partial t}+\frac{\partial\rho}{\partial t}+\overrightarrow{\nabla}.\overrightarrow{j}-\overrightarrow{\nabla}.\overrightarrow{k} & \frac{\partial\overrightarrow{j}}{\partial t}+\frac{\partial\overrightarrow{k}}{\partial t}+\overrightarrow{\nabla}\varrho+\overrightarrow{\nabla}\rho-\overrightarrow{\nabla}\times\overrightarrow{\overrightarrow{j}}+\overrightarrow{\nabla}\times\overrightarrow{k}\\
-\overrightarrow{\nabla}\varrho+\overrightarrow{\nabla}\rho+\frac{\partial\overrightarrow{j}}{\partial t}-\frac{\partial\overrightarrow{k}}{\partial t}+\overrightarrow{\nabla}\times\overrightarrow{\overrightarrow{j}}+\overrightarrow{\nabla}\times\overrightarrow{k} & \overrightarrow{\nabla}.\overrightarrow{j}+\overrightarrow{\nabla}.\overrightarrow{k}+\frac{\partial\varrho}{\partial t}+\frac{\partial\rho}{\partial t}\end{array}\right)\label{eq:28}\end{align}
which can be written as 

\begin{align}
\overline{\boxdot}\mathcal{\mathbb{J}}= & \mathbb{S}\label{eq:29}\end{align}
 where

\begin{align}
\mathbb{S}= & \left(\begin{array}{cc}
\left(\Im_{m}-\Im_{e}\right) & -\left(\overrightarrow{r}+\overrightarrow{s}\right)\\
\left(\overrightarrow{r}-\overrightarrow{s}\right) & \left(\Im_{m}+\Im_{e}\right)\end{array}\right)\longmapsto\left(\begin{array}{cc}
\Im_{-} & \overrightarrow{S_{+}}\\
\overrightarrow{S_{-}} & \Im_{+}\end{array}\right)\label{eq:30}\end{align}
with $\begin{array}{c}
\Im_{-}\end{array}=\left(\Im_{m}-\Im_{e}\right),\;\overrightarrow{S_{-}}=\left(\overrightarrow{r}-\overrightarrow{s}\right),\;\Im_{+}=\left(\Im_{m}+\Im_{e}\right),\;\overrightarrow{S_{+}}=\left(\overrightarrow{r}+\overrightarrow{s}\right)$ and 

\begin{align}
\Im_{m}= & \overrightarrow{\nabla}.\overrightarrow{k}+\frac{\partial\varrho}{\partial t};\,\,\,\,\,\,\,\Im_{e}=\overrightarrow{\nabla}.\overrightarrow{j}+\frac{\partial\rho}{\partial t}\label{eq:31}\end{align}
 along with \begin{align}
\overrightarrow{r}= & -\overrightarrow{\nabla}\rho-\frac{\partial\overrightarrow{j}}{\partial t}-\overrightarrow{\nabla}\times\overrightarrow{k};\,\,\,\,\,\overrightarrow{s}=-\overrightarrow{\nabla}\varrho-\frac{\partial\overrightarrow{k}}{\partial t}+\overrightarrow{\nabla}\times\overrightarrow{j}\label{eq:32}\end{align}
which are the split octonion current wave equations for the components
of generalized fields of dyons. In equation (\ref{eq:30}) $\Im_{m}$
and $\Im_{e}$are vanishing due to Lorentz gauge conditions applied
for the cases of electric and magnetic charges.

As such, we have obtained consistently the generalized Dirac Maxwell's
(GDM) equations from the theory of split octonion variables without
constraints. The advantages of present formalism are discussed in
terms of compact and simpler notations of split octonion valued generalized
potential, generalized field and generalized currents of dyons. The
present split octonion reformulation of generalized fields of dyons
represents well the invariance of field equations under Lorentz and
duality transformations. It also reproduces the dynamics of electric
(magnetic) charge yielding to the usual form of Maxwell's equations
in the absence of magnetic (electric charge) in compact, simpler and
consistent way. 

\textbf{Acknowledgment: }One of us OPSN is thankful to Professor H.
Dehnen, Universität Konstanz, Fachbereich Physik, Postfach-M 677,
D-78457 Konstanz, Germany for his kind hospitality at Universität
Konstanz. He is also grateful to German Academic Exchange Service
(Deutscher Akademischer Austausch Dienst), Bonn for their financial
support under DAAD re-invitation programme.

\end{document}